\documentstyle[12pt]{article}

\begin{document}
\begin{titlepage}
\begin{center}

January 12, 2001     \hfill    LBNL-44712 \\

\vskip .2in

{\large \bf Von Neumann's Formulation of Quantum Theory and the Role
of Mind in Nature}
\footnote{This work is supported in part by the Director, Office of Science, 
Office of High Energy and Nuclear Physics, Division of High Energy Physics, 
of the U.S. Department of Energy under Contract DE-AC03-76SF00098}

\vskip .20in
Henry P. Stapp\\
{\em Lawrence Berkeley National Laboratory\\
      University of California\\
    Berkeley, California 94720}
\end{center}

\vskip .2in

\begin{abstract}
Orthodox Copenhagen quantum theory renounces the quest to understand the 
reality in which we are imbedded, and settles for practical rules that 
describe connections between our observations. Many physicist have believed
that this renunciation of the attempt describe nature herself was premature,
and John von Neumann, in a major work, reformulated quantum theory as 
theory of the evolving objective universe. In the course of his work
he converted to a benefit what had appeared to be a severe deficiency of 
the Copenhagen interpretation, namely its introduction into physical theory 
of the human observers. He used this subjective element of quantum  theory to
achieve a significant advance on the main problem in philosophy, which
is to understand the relationship between mind and matter. That problem
had been tied closely to physical theory by the works of Newton and
Descartes. The present work examines the major problems that have
appeared to block the development of von Neumann's theory into a 
fully satisfactory theory of Nature, and proposes solutions to these
problems.

\end{abstract}

\end{titlepage}

\newpage
\renewcommand{\thepage}{\arabic{page}}
\setcounter{page}{1}

{\bf The Nonlocality Controversy}

``Nonlocality gets more real''. This is the provocative title of a recent 
report in Physics Today [1]. Three experiments are cited. All three
confirm to high accuracy the predictions of quantum theory in experiments 
that suggest the occurrence of an instantaneous action over a large distance. 
The most spectacular of the three experiments begins with the 
production of pairs of photons in a lab in downtown Geneva. For some of
these pairs, one member is sent by optical fiber to the village 
of Bellevue, while the other is sent to the town of Bernex. 
The two towns lie more than 10 kilometers apart. Experiments on the arriving 
photons are performed in both villages at essentially the same time. What 
is found is this: The observed connections between the outcomes of these 
experiments defy explanation in terms of ordinary ideas about the nature 
of the physical world {\it on the scale of directly observable objects.}
This conclusion is announced in opening sentence of the 
Physical-Review-Letters report [2] that describes the experiment: 
``Quantum theory is nonlocal''. 

This observed effect is not just an academic matter. A possible application
of interest to the Swiss is this: The effect can be used in principle to 
transfer banking records over large distances in a secure way [3]. 
But of far greater importance to physicists is its relevance to two 
fundamental questions: What is the nature of physical reality? What is the 
form of basic physical theory? 

The answers to these questions depend crucially on the nature of physical 
causation. Isaac Newton erected  his theory of gravity on the idea of instant 
action at a distance. According to Newton's theory, if a person were to
suddenly kick a stone, and send it flying off in some direction, every 
particle in the entire universe would {\it immediately} begin to feel the 
effect of that kick. Thus, in Newton's theory, every part of the universe is 
instantly linked, causally, to every other part. To even think about such an 
instantaneous action one needs the idea of the instant of time ``now'',
and a sequence of such instants each extending over the entire universe.

This idea that what a person does in one place  could act instantly affect
physical reality in a faraway place is a mind-boggling notion, and it was 
banished from classical physics by Einstein's theory of relativity. But the 
idea resurfaced at the quantum level in the debate between Einstein and Bohr. 
Einstein objected to the ``mysterious action at a distance'', which quantum 
theory seemed to entail, but Bohr defended  ``the necessity of a final 
renunciation of the classical ideal of causality and a radical revision of 
our attitude towards the problem of physical reality''[4]. 

The essence of this radical revision was explained by Dirac at the 1927 Solvay
conference [5]. He insisted on the restriction of the application of quantum 
theory to our knowledge of a system, rather than to the system itself. 
Thus physical theory became converted from a theory about `physically reality',
as it had formerly been understood, into a theory about human knowledge.

This view is encapsulated in Heisenberg's famous statement [6]:

``The conception of the objective reality of the elementary particles
has thus evaporated not into the cloud of some obscure new reality
concept, but into the transparent clarity of a mathematics that represents
no longer the behaviour of the particle but rather our knowledge of this
behaviour.''  

This conception of quantum theory, espoused by Bohr, Dirac, and Heisenberg,
is called the Copenhagen interpretation. It is essentially
subjective and epistemological, because the basic reality of the
theory is `our knowledge'.

It is certainly true that science rests ultimately on what we know.
That fact is the basis of the new point of view. However,
the tremendous successes of the classical physical theory inaugurated
by Galileo, Descartes, and Newton during the seventeenth century,
had raised the hope and expectation that human beings could extract from 
careful observation, and the imaginative creation  of testable hypotheses, 
a valid idea of the general nature and rules of behaviour of 
the reality in which our human knowledge is imbedded. Giving up on that
hope is indeed a radical shift. On the other hand, classical physical 
theory left part of reality out, namely our conscious experiences. 
Hence it had no way to account either for the existence of our conscious 
experiences or for how knowledge can reside in those
experiences. Thus bringing human experience into our understanding of reality
seems to be a step in the right direction. It might allow science to explain, 
eventually, how we know what we know. But Copenhagen quantum theory is only 
a half-way house: it brings in human experience, but at the stiff price of 
excluding the rest of reality. 
 
Yet how could the renowned scientists who created Copenhagen quantum theory 
ever believe, and sway most other physicists into believing, that a 
complete science could leave out the physical world? It is undeniable that
we can never know for sure that a proposed theory of the world around us is 
really true. But that is not a sufficient reason to renounce, as a matter of 
principle, the attempt to form at least a coherent idea of what the world 
{\it could} be like, and rules by which it {\it could} work. Clearly some 
extraordinarily powerful consideration was in play.  

That powerful consideration was a basic idea about the nature of physical
causation that had been injected into physics by Einstein's theory 
of relativity. That idea was not working!

The problem is this.
Quantum theory often entails that an act of acquiring knowledge in one place 
instantly changes the theoretical representation of some faraway system. 
Physicists were---and are---reluctant to believe that performing a nearby 
act can instantly change a faraway physical reality. However, they recognize
that ``our knowledge'' of a faraway system can instantly change when we learn  
something about a nearby system. In particular, if certain properties of 
two systems are known to be strongly correlated, then finding out something
about one system can tell us something about the other. For example, 
if we know that two particles start from some known point at the same time, 
and then move away from that point at the same speeds, but in opposite 
directions, then finding one of these particles at a certain point allows 
us to `know' where the other particle lies at that same instant: it must 
lie at the same distance from the starting point as the observed particle,
but in the opposite direction. In this simple case we do not think that 
the act of observing the position of one particle {\it causes} the other 
particle to {\it be} where it is. We realize that it is only our knowledge 
of the faraway system that has changed. This analogy allows us resolve, 
by fiat, any mystery about an instantaneous faraway effect of a nearby act: 
if something faraway can instantly be altered by a nearby act then it 
{\it must be} our knowledge. But then the analog in quantum theory of the 
physical reality of classical physical theory {\it must be} our knowledge.

This way of dodging the action-at-a-distance problem was challenged by 
Einstein, Podolsky, and Rosen in a famous paper [7] entitled: ``Can 
quantum-mechanical description of physical reality be considered complete?'' 
The issue was whether a theory that is specified to be merely a set of rules 
about connections between human experiences can be considered to be a complete 
description of physical reality. Einstein and his colleagues gave a 
reasonable definition of ``physical reality'', and then argued, directly 
from some basic precepts of quantum theory itself, that the answer to this 
question is `No'. Bohr [8] disagreed. 

Given the enormity of what must exist in the universe as a whole, and the 
relative smallness human knowledge, it is astonishing that, in the minds of 
most physicists, Bohr prevailed over Einstein in this debate: the majority
of quantum physicists acquiesced to Bohr's claim that 
quantum theory, regarded as a theory about human knowledge, is a 
complete description of physical reality. This majority opinion stems, 
I believe, more from the lack of a promising alternative candidate 
than from any decisive logical argument.

Einstein, commenting on the orthodox Copenhagen position, said:  
``What I dislike about this kind of argument is the basic positivistic 
attitude,  which from my view is untenable, and seems to me to come to 
the same thing as Berkeley's principle [9], {\it esse est percipi}, ``to be 
is to be perceived''. Several other scientists also reject the majority 
opinion. For example, Murray  Gell-Mann [10] asserts: ``Niels  Bohr 
brainwashed a whole generation into believing that the problem was solved 
fifty years ago''. Gell-mann believes that in order to integrate 
quantum theory coherently into cosmology, and to understand the evolutionary 
process that has  produced creatures that can have knowledge, one needs to 
have a coherent theory of the evolving quantum mechanical reality in which 
these creatures are imbedded.

It is in the context of such efforts to construct a more complete theory
that the significance of the experiments pertaining to quantum nonlocality
lies.

The point is this: If nature really is nonlocal, as these experiments
suggest, then the way is open to the development of a rationally coherent 
theory of nature that integrates the subjective knowings introduced by 
Copenhagen quantum theory into an objectively existing and evolving
physical reality. The basic framework is provided by the version of quantum 
theory constructed by John von Neumann [11] 

All physical theories are, of course, provisional, and subject to future 
revision and elaboration. But at a given stage in the development of science 
the contending theories can be evaluated on many grounds, such as utility, 
parsimony, predictive power, explanatory power, conceptual simplicity, 
logical coherence, and aesthetic beauty. The development of von Neumann's
theory that I shall describe here fares well on all of these counts. 

To understand von Neumann's improvement one must appreciate the
problems with its predecessor. Copenhagen quantum theory gives special 
status to measuring devices. These devices are physical systems: they are 
made up of atomic constituents. But in spite of this, these devices are 
excluded from the world of atomic constituents that are described in the 
mathematical language of quantum theory. The measuring devices, are described,
instead, in a different language, namely by ``the same means of communication 
as the one used in classical physics'' [12]. This approach renders the theory 
pragmatically useful but physically incoherent. It links the theory to 
``our knowledge'' of the measuring devices in a useful way, but disrupts
the dynamical unity of the physical world by treating in different ways
different  atomic particles that are interacting with each other. This tearing 
apart of the physical world creates huge conceptual problems, which are ducked 
in the Copenhagen approach by renouncing man's ability to understand reality. 

The Copenhagen version of quantum theory is thus a hybrid of the old 
familiar classical theory, which physicists were understandably reluctant 
to abandon completely, and a totally new theory based on radically 
different concepts. The old ideas, concepts, and language were used to
describe our experiences, but the old idea that the visible objects
were made up of tiny material objects resembling miniature planets, 
or minute rocks, was dropped. The observed physical world is described 
rather by a mathematical structure that can best be characterized 
as representing {\it information} and {\it propensities}: the 
{\it information} is about certain {\it events} that have occurred in the 
past, and the {\it propensities} are objective tendencies pertaining to future 
events. 

These ``events'' are the focal point of quantum theory: they are 
happenings that in the Copenhagen approach are ambiguously
associated both with the ``measuring devices'' and with increments in the
knowledge of the observers who are examining these devices. Each increment 
of knowledge is an event that updates the knowledge of the observers by 
bringing it in line with the observed outcome of an event occurring at 
a device. The agreement between the event at the device and the event 
in the mind of the observer is to be understood in the same way 
as it is understood in classical physics.

But there's the rub: the connection between human knowledge and the
physical world never has been understood in classical physics.
The seventeenth century division between mind and matter upon which 
classical physically theory was erected was such a perfect cleavage that 
no reconciliation has ever been achieved, in spite of tremendous efforts. 
Nor is such a reconciliation possible within classical 
physics. According to that theory, the world of matter 
is built out of
microscopic entities whose behaviours are fixed by interaction
with their immediate neighbors. Nothing need exist except
what can be deduced, by using only the precepts
of classical physical theory, from the existence of these
microscopic building blocks. But the defining characteristic
of consciousness, namely its experiential quality,  not deducible
from these elements of classical physical theory, and that is all 
that classical physical theory entails.

The fact that quantum theory is intrinsically a theory of mind-matter
interaction was not lost upon the early founders and workers. Wolfgang Pauli,
John von Neumann, and Eugene Wigner were three of the most rigorous thinkers 
of that time. They all recognized that quantum theory
was about the mind-brain connection, and they tried to develop that idea. 
However, most physicists were more interested in experiments on
relatively simple atomic systems, and were understandably reluctant to 
get sucked into the huge question of the connection between mind and brain. 
They were willing to sacrifice certain formerly-held ideals, and take
practical success to be the basic criterion of  good science. 

This retreat
both buttressed, and was buttressed by, two of the main philosophical 
movements of the twentieth century. One of these, 
materialism-behaviourism, effectively denied the existence 
of our conscious ``inner lives'', and the other, 
postmodern-social-consructionism, viewed science as a social
construct without any objective mind-independent content. The 
time was not yet ripe, either philosophically or scientifically,  for a serious attempt to
study the physics of mind-matter connection. Today, however,
as we enter the third millenium, there is a huge surge 
of interest among
philosophers, psychologists, and neuroscientists in 
reconnecting the aspects of nature that were torn asunder 
by seventeenth century physicists.

John von Neumann was one of the most brilliant mathematicians and logicians 
of his age. He followed where the mathematics and logic led. 
From the point of view of the mathematics of quantum theory it makes no 
sense to treat a measuring device as intrisically different from the 
collection of atomic constituents that make it up. A device is just another
 part of the physical universe, and it should be treated as such. Moreover, 
the conscious thoughts of a human observer ought to be causally connected 
{\it most directly and immediately} to what is happening in his brain, 
not to what is happening out at some measuring device. 

The mathematical rules of 
quantum theory specify clearly how the measuring devices are to be included 
in the quantum mechanically described physical world. Von Neumann first 
formulated carefully the mathematical rules of quantum theory, and then 
followed where that mathematics led. It led first to the incorporation of 
the measuring devices into the quantum mechanically described physical 
universe, and eventually to the inclusion of {\it everything} built out of 
atoms and their constituents. Our bodies and brains thus become, in 
von Neumann's approach, parts of the quantum mechanically described physical 
universe. Treating the entire physical universe in this unified way provides 
a conceptually simple and logically coherent theoretical foundation that
heals the rupturing of the physical world introduced by the Copenhagen
approach. It postulates, for each observer, that each experiential
event is connected in a certain specified way to a corresponding brain event. 
The dynamical rules that connect mind and brain are very restrictive, and this
leads to a mind-brain theory with significant explanatory 
power.

Von Neumann showed in principle how all of the predictions of 
Copenhagen quantum theory are contained in his version.
However, von Neumann quantum theory gives, in principle, much 
more than Copenhagen quantum theory can. By providing an 
objective description of the entire history of the universe, 
rather than merely rules connecting human observations, von 
Neumann's theory provides a quantum framework for
cosmological and biological evolution. And by including both 
brain and knowledge, and also the dynamical laws that connect them, 
the theory provides a rationally coherent dynamical framework for 
understanding the relationship between brain and mind.

There is, however, one major obstacle: von Neumann's theory,
as he formulated it, appears to conflict with Einstein's 
theory of relativity.

{\bf Reconciliation with Relativity}

Von Neumann formulated his theory in a nonrelativistic approximation:
he made no attempt to reconcile it with the empirically validated features 
of Einstein's theory of relativity.

Today this reconciliation is easily achieved. One can simply replace 
the nonrelativistic theory used by von Neumann with relativistic 
quantum field theory. 

To deal  with 
the mind-brain interaction one needs to consider the physical processes in 
human brains. The relevant quantum field theory is quantum-elecrodynamics. 
The relevant energy range is that of atomic and molecular interactions. 
I shall assume that whatever high-energy theory eventually prevails in 
quantum physics, it will reduce to quantum electrodynamics in this 
low-energy regime.

But there remains one apparent problem: von Neumann's nonrelativistic theory 
is built on the Newtonian concept of the instants of time, `now', each of 
which extends over all space. The evolving state of the universe, $S(t)$, is 
defined to be the state of the entire universe at the instant of time t. 
However, Einstein's theory of relativity rejected, at least within classical
physical theory, the idea that the Newtonian concept of the instant ``now''
could have any objective meaning.

Tomonaga [13] and Schwinger [14] have constructed a standard formulation of 
relativistic quantum field theory. It has effective instants ``now'', namely 
the Tomonaga-Schwinger surfaces $\sigma $. Pauli once strongly emphasized 
to me that these surfaces, while they give a certain aura of relativistic 
invariance, do not differ significantly from the constant-time surfaces 
``now'' that appear in the Newtonian physics. All efforts to remove 
completely from quantum theory the distinctive character of time, 
in comparison to space, have failed.

To obtain a relativistic version of von Neumann's theory one 
need merely identify the sequence of constant-time surfaces ``now'' in 
his theory with a corresponding objectively defined sequence 
of Tomonaga-Schwinger surfaces $\sigma $.

Giving special objective physical status to a particular sequence of 
spacelike surfaces $\sigma$, say the constant-time surfaces in some one 
particular Lorentz  frame, does not disrupt the covariance properties
of the {\it empirical predictions}  of the theory: that was one of the 
main consequences of the Tomonoga-Schwinger formulation. Although each
reduction of the state vector is taken to occur instantaneously 
along one of the preferred set of surfaces $\sigma$  the predictions
about observations remain independent of {\it which} sequence of
surfaces $\sigma$ is chosen (e.g., which Lorentz frame is used
to define the preferred sequence of surfaces). Thus this relativistic 
version of von Neumann's theory is fully compatible with the theory of 
relativity at the level of empirically accessible relationships. However,
the theory does conflict with a {\it metaphysical idea} spawned by the theory 
of relativity, namely the idea that there is no dynamically preferred 
sequence of instantaneous ``nows''. Thus theory reverts, at a certain
deep ontological level, to the Newtonian idea of instantaneous action at 
a distance, while maintaining all of the empirical demands of the theory of 
relativity.

The astronomical data [15] indicates that there 
does exist, in the observed universe, a preferred sequence 
of `nows': they define the special set of  surfaces in which, for the 
early universe, matter was distributed almost uniformly in mean local 
velocity, temperature, and density. It is natural to assume that these
empirically specified surfaces are the same as the objective preferred
surfaces ``now'' of von Neumann quantum theory.

{\bf Nonlocality and Relativity}

von Neumann's objective theory immediately accounts for the faster-than-light 
transfer of information that seems to be entailed by the nonlocality 
experiments: the outcome that appears first, in the cited experiment, 
occurs in one or the other of the two Swiss villages. According to the theory,
this earlier event has an immediate instantaneous effect on the evolving 
state of the universe, and this change has an immediate effect on the 
{\it propensities} for the various possible outcomes of the measurement 
performed slightly later in the other village. 

This feature---that there is some sort of objective instantaneous transfer
of information---conflicts with the spirit of the theory of relativity. 
However, this  quantum effect is of a subtle kind: it acts neither on 
material substance, nor on locally conserved energy-momentum, nor on anything 
else that exists in the classical conception of the physical world that the 
theory of relativity was originally designed to cover. It acts on a 
mathematical structure that represents, rather, {\it information and 
propensities}.

The theory of relativity was originally formulated within classical physical
theory. This is a deterministic theory: the entire history of the universe 
is completely determined by how things started out. Hence all of history 
can be conceived to be laid out in a four-dimensional spacetime. The idea of 
``becoming'', or of the gradual unfolding of reality, has no natural place in
this deterministic conception of the universe. 

Quantum theory is a different kind of theory: it is formulated
as an indeterministic theory. Determinism is relaxed in two important
ways. First, freedom is granted to each experimenter to choose freely
which experiment he will perform, i.e., which aspect of nature
he will probe; which question he will put to nature. Then Nature is allowed 
to pick an outcome of the experiment, i.e., to answer to the question. This
answer is partially free: it is subject only to certain statistical 
requirements. These elements of `freedom of choice', on the part of both 
the human participant and Nature herself, lead to a picture of a reality 
that gradually unfolds in response to choices that are not necessarily 
fixed by the prior physical part of reality alone. 

The central roles in quantum theory of these discrete choices--- 
the choices of which questions will be put to nature, and which answer 
nature delivers---makes quantum theory a theory of discrete events, rather 
than a theory of the continuous evolution of locally conserved matter/energy. 
The basic building blocks of the new conception of nature are not objective 
tiny bits of matter, but choices of questions and answers.
 
In view of these deep structural differences there is a question of 
principle regarding how the stipulation that there can be no faster-than-light 
transfer of information of any kind should be carried over from the invalid 
deterministic classical theory to its indeterministic quantum successor.

The theoretical advantages of relaxing this condition are great: it provides
an immediate resolution all of the causality puzzles that have blocked 
attempts to understand physical reality, and that have led directly to the
Copenhagen renunciation of all such efforts. And it hands to us a new rational 
theoretical basis for attacking the age-old problem of the connection 
between mind and brain.

In view of these potential advantages one must ask whether
it is really beneficial for scientists to renounce for all 
time the aim of trying to understand the world in which we live, 
in order to maintain a metaphysical prejudice that arose from a
theory that is known to be fundamentally incorrect?

I use the term  ``metaphysical prejudice'' because there is no theoretical
or empirical evidence that supports the non-existence of the subtle sort
of instantaneous action that is involved here. Indeed, both theory and the 
nonlocality experiments, taken at face value, seem to demand it. 
The denial of the possibility of any such action is a metaphysical
commitment that was  useful in the context of classical physical theory. 
But that earlier theory contains no counterpart of the informational structure 
upon which the putative action in question acts.

Renouncing the endeavour to understand nature is a price too heavy to pay 
to preserve a metaphysical prejudice.

{\bf Is Nonlocality Real?}

I began this article with the quote from Physics Today: ``Nonlocality
gets more real.'' The article described experiments whose outcomes were 
interpreted as empirical evidence that nature was nonlocal, in some
sense. But do nonlocality experiments of this kind provide any real 
evidence that information is actually transferred over spacelike intervals? 
An affirmative answer to this question would provide
support for rejecting the metaphysical prejudice in question 

The evidence is very strong that the predictions  of quantum theory
are valid in these experiments involving pairs of measurements
performed at essentially the same time in regions lying far apart.
But the question is this: Does the fact that the predictions of 
quantum theory are correct in experiments of this kind
actually show that information must be transferred instantaneously, 
in some (Lorentz) frame of reference?

The usual arguments that connect these experiments to nonlocal action 
stem from the work of John Bell [16]. What Bell did was this. He noted
that the argument of Einstein, Podolsky, and Rosen was based on a certain
assumption, namely that ``Physical Reality'', whatever it was, should
have at least one key property: What is physically real in one region
should not depend upon which experiment an experimenter in a faraway region
freely chooses to do at essentially the same instant of time. Einstein
and his collaborators showed that if this property is valid then the 
physical reality in a certain region must include, or specify, the values 
that certain unperformed measurements {\it would have revealed} if they,
rather than the actually performed measurements, had been performed. However, 
Copenhagen quantum theory cannot accommodate well defined outcomes of these
not-actually-performed measurements. Thus the Einstein-Podilsky-Rosen 
argument, if correct, would prove that the quantum framework cannot be a
complete description of physical reality.  

Bohr countered this argument by rejecting the claimed key property 
of physical reality: he denied the claim pertaining to no instantaneous 
action at a distance. That rebuttal is subtle, and many physicists 
(e.g. Einstein and John Bell) and philosophers (e.g. Karl Popper) doubted
that Bohr had successfully answered the EPR argument.

Bell found a more direct way to counter the argument of  Einstein, Podolsky, 
and Rosen. He accepted both their claim that the results of these
{it unperformed measurements} are indeed physically real, and that these 
physical realities could not be influenced by what far-away experimenters 
choose to measure at essentially the same instant of time. And he assumed, 
as did all the disputants, that the {\it predictions} of quantum theory were 
correct. 

From these assumptions Bell deduced a mathematical contradiction, thereby 
showing that {\it something} must be wrong with either the conclusions of 
Einstein, Podolsky, and Rosen, or with the no-faster-than-light-influence 
assumption. But Bell's argument did not fixed exactly where the trouble lies. 
Does the trouble lie with the assumption of no faster-than-light influence, 
or with the EPR conclusion that the outcomes of the unperformed measurements 
are physically real?

Orthodox quantum theorists have no trouble answering this question:
the {\it assumption} that outcomes of unperformed measurements are physically
real is wrong. This idea  directly contradicts quantum philosophy! 

This answer allows one to retain Einstein's reasonable-sounding assumption 
that physical reality in one place cannot be influence by what a faraway 
experimenter freely chooses to do at the same instant: the Bell's 
argument neither entails, nor even suggests, the existence of 
faster-than-light influences. 

Bell, and others who followed his ``hidden-variable'' approach,
later used assumptions that appear weaker than this original 
one. However, these later assumptions are essentially the same
as the earlier one: they turn out to entail [17, 18] the possibility 
of defining numbers that could specify, simultaneously, the values 
that all the relevant unperformed measurements would  reveal if they were 
to be performed. But, as just mentioned, one of the basic 
precepts quantum philosophy is that such numbers
do not exist.

{\bf Eliminating Hidden Variables}

The purpose of Bell's argument is different from that of Einstein,
Podolsky, and Rosen, and the logical demands are different. The challenge 
faced by Einstein and his colleagues was to mount an argument built directly 
on the orthodox quantum principles themselves. For only by proceeding in this 
way could they get a logical hook on the quantum physicists that they 
wanted to convince. 

This demand posed a serious problem for Einstein and co-workers. Their 
argument, like Bell's, involved a consideration of the values that 
unperformed measurements would reveal if they were to be performed. Indeed, 
it was precisely the Copenhagen claim that such values do not exist that 
Einstein and company wanted to prove untenable. But they needed to establish 
the existence of such values without begging the question by making an 
assumption that was equivalent to what the were trying to show.

The strategy of Einstein et. al. was to prove the existence of such
values by using only quantum precepts themselves, plus the seemingly 
secure idea from the theory of relativity that what is physically real 
here and now cannot be influenced by what a faraway 
experimenter chooses to do now.

This strategy succeeded: Bohr was forced into an awkward position of
rejecting Einstein's premise that ``physical reality'' could not be influenced 
by what a faraway experimenter chooses to do:

``...there is essentally the question of {\it an influence on the very 
conditions which define the possible types of predictions regarding future 
behavior of the system.} Since these conditions constitute an inherent 
element of any phenomena to which the term `physically reality' can be 
properly attached we see that the argument of 
mentioned authors does not justify their conclusion that quantum-mechanical
description is essentially incomplete.''[8]

I shall pursue here a strategy similar to that of Einstein and his colleagues,
and will be led to a conclusion similar to Bohr's, namely the failure
of Einstein's assumption that physical reality cannot be influenced from 
afar.

Values of unperformed measurements can be brought into the theoretical 
analysis by combining two ideas that are embraced by Copenhagen philosophy. 
The first of these is the freedom of experimenters to choose which 
measurements they will perform. In Bohr's words:

``The freedom of experimentation, presupposed in classical physics,
is of course retained and corresponds to the free choice of experimental
arrangements for which the mathematical structure of the quantum
mechanical formalism offers the appropriate latitude.''[19]

This assumption is important for Bohr's notion of complementarity:
some information about all the possible choices is simultaneously present
in the quantum state, and Bohr wanted to provide the possibility that
any one of the mutually exclusive alternatives might be pertinent. Whichever
choice the experimenter eventually makes, the associated set of  
predictions is assumed to hold.

The second idea is the condition of no backward-in-time
causation. According to quantum thinking, experimenters are to be considered 
free to choose which measurement they will perform. Moreover, if an outcome 
of a measurement appears to an observer at a time earlier than some time $T$, 
then this outcome can be considered to be fixed and settled at that time $T$, 
independently of which experiment will be {\it freely chosen} and performed 
by another experimenter at a time later than $T$: the later choice is allowed 
go either way without disturbing the outcome that has already appeared  
to observers at an earlier time.

I shall make the {\it weak} assumption that this no-backward-in-time-influence 
condition holds for {\it at least one} coordinate system (x,y,z,t).

These two conditions are, I believe, completely compatible with quantum 
thinking, and are a normal part of orthodox quantum thinking. They
contradict no quantum precept or combination of  quantum predictions. 
They, by themselves, lead to no contradiction. But they do introduce
into the theoretical framework a very limited notion of a result of 
an unperformed measurement, namely the result of a measurement that is  
actually performed in one region at an earlier time $T$ coupled with the 
measurment NOT performed {\it later} by some faraway experimenter. My 
assumption is that this early outcome, which is actually observered by
someone, can be treated as existing independently of which of the two 
alternative choices is made by the experimenter in the later region, 
even though only one of the two later options can be realized. This 
assumption of no influence backward in time constitutes the small element of 
counterfactuality that provides the needed logical toe-hold.

{\bf The Hardy Experimental Setup}

To get a nonlocality conclusion like the one obtained from  Bell-type
theorems, but from assumptions that are in line with the 
precepts  of quantum theory,
it is easiest to consider an experiment of the kind first  discussed
by Lucien Hardy [20]. The setup is basically similar to the ones considered 
in proofs of Bell's theorem. There are two spacetime regions, L and R, that 
are ``spacelike separated''. This condition means that the two regions are 
situated far apart in space relative to their extensions in time, so that 
no point in either region can be reached from any point in the other 
without moving either faster than the speed of light or backward in time.
This means also that in some frame of reference, which I take to be the 
coordinate system (x,y,z,t) mentioned above, the region L lies at times 
greater than time $T$, and the region  R lies earlier than time $T$.   

In each region an experimenter freely chooses between two possible
experiments. Each experiment will, if chosen, be performed within that region,
and its outcome will appear to observers within that region.
Thus neither choice can affect anything located in the other region without 
there being some influence that acts faster than the speed of light 
or backward in time.

The argument involves four predictions made by quantum theory
under the Hardy conditions. These conditions and predictions 
are described in Box 1.

--------------------------------------------------------------------

{\bf Box 1: Predictions of quantum theory for the Hardy experiment.}

The two possible experiments in region  L are labelled L1 and L2.

The two possible experiments in region  R are labelled R1 and R2.

The two possible outcomes of L1 are labelled L1+ and L1-, etc.

The Hardy setup involves a laser down-conversion source that emits a pair 
of correlated photons. The experimental conditions are such that 
quantum theory makes the following four  predictions:\\ \\
1. If (L1,R2) is performed and L1- appears in L then R2+ must appear in R.\\
2. If (L2,R2) is performed and R2+ appears in R then L2+ must appear in L.\\
3. If (L2,R1) is performed and L2+ appears in L then R1- must appear in R.\\
4. If (L1,R1) is performed and L1- appears in L then R1+ appears sometimes
   in R.\\ 

The three words ``must'' mean that the specified outcome is predicted
to occur with certainty (i.e., probability unity).\\ 
---------------------------------------------------------------------------

{\bf Two Simple Conclusions}

It is easy to deduce from our assumptions two simple conclusions.

Recall that region R lies earlier than time $T$, and that region L lies
later than time $T$.

Suppose the actually selected pair of experiments is (R2, L1), and
that the outcome L1- appears in region L. Then prediction 1 of 
quantum theory entails that R2+ must have already appeared in R prior to time
$T$. The no-backward-in-time-influence condition then entails that this 
outcome R2+ was fixed and settled prior to time $T$, independently of 
which way the later free choice in L will eventually go: the outcome in region
R at the earlier time would still be R2+ even if the later free choice 
had gone the other way, and L2 had been chosen {\it instead of} L1.

Under this alternative condition (L2,R2,R2+) the experiment L1 would not 
be performed, and there would be no physical reality corresponding to its 
outcome. But the actual outcome in R would still be R2+, and we are assuming 
that the predictions of quantum theory will hold no matter which of the two
experiments is eventually performed later in L. Prediction 2 of quantum 
theory asserts that it must be  L2+. This yields the following conclusion: 

Assertion A(R2):

If (R2,L1) is performed and outcome L1- appears in region L, then if
the choice in L had gone the other way, and L2, instead of L1, had been
performed in L then outcome L2+ would have appeared there.  

Because we have two predictions that hold with certainty, and the two
strong assumptions of `free choice' and `no backward causation', it is 
not surprising that we have been able to derive this conclusion. In an 
essentially deterministic context we are often able to deduce from
the outcome of one measurement what would have happened if we had made,
instead, another measurement. Indeed, if knowing the later {\it actual} 
outcome allows one to know what some earlier condition must have been, and 
if this earlier condition entails a unique result of the later 
{\it alternative} measurement, then one can conclude from knowledge
of the later {\it actual} outcome what would have happened if, instead, the 
later {\it alternative} measurement had been performed. This is about 
the simplest possible example of counterfactual reasoning.  

Consider next the same assertion, but with R2 replaced by R1:

Assertion A(R1):

If (R1,L1) is performed and outcome L1- appears in region L, then if 
the choice in L had gone the other way, and L2, instead of L1, had been
performed in L then outcome L2+ would have appeared there.  

This assertion cannot be true. The fourth prediction of quantum theory asserts 
that under the specified conditions L1- and R1 the outcome R1+ sometimes 
appears in R. The no-backward-in-time-influence condition ensures that this
earlier appearance of R1+ would not be altered if the later choice in 
region L had gone the other way and L2 had been chosen there: that is our
basic causality assumption. But A(R1) asserts that under this condition that 
L2 is performed later the outcome L2+ must appear in  L. But if L2+
were to occur under this condition (L2,R1) then the third prediction entails 
that then R1- must appear in R. That conclusion contradicts the previously 
established result that under these conditions R1+ sometimes appears in R.

Thus, given the validity of the four predictions of quantum theory, 
and of our basic causality condition, the validity of assertion A(R1) 
cannot be maintained.

The fact that A(R2) is true and A(R1) is false entails a certain nonlocal
connection. The truth of the first of these statements means that certain 
assumptions that are {\it compatible} with quantum philosophy, and that 
probably are {\it parts} of quantum philosophy, as it is understood by most 
quantum physicists, entail a connection of the following form: 
``If experiment  E1 is  performed in region L and gives outcome O1 
in region L then if, instead, experiment E2 had been performed in region L
the outcome in region L  would have been O2.'' This result is similar to 
what Einstein, Podolsky, and Rosen tried to prove, but is weaker,
because it does not claim that the two outcomes exist simultaneously. However, 
it is derived from weaker assumptions: it is not based on the criterion 
for ``Physical Reality'' that Einstein, Podolsky, and Rosen used, but 
Bohr rejected. This weaker conclusion, alone, does not directly contradict any 
precept of quantum philosophy. But the {\it conjunction} of 
the two statements, ``A(R2) is true and A(R1) is false'' leads to a
problem. It asserts that a theoretical constraint upon what nature can
choose in region L, under conditions freely chosen by the experimenter 
in region L, depends nontrivially on which experiment is freely
chosen by the experimenter in regions R: a theoretical constraint on Nature's
choices in L depends upon what a faraway experimenter freely decides to
do in R. Any theoretical model that is compatible with the premises 
of the argument would have to maintain these theoretical constraints on 
nature's choices in region L, and hence enforce the nontrivial dependence
of these constraints on the free choice made in region R. But this
dependence cannot be upheld without the information about the 
free choice made in region R getting to region L: {\it some sort of 
faster-than-light tranfer of information is required.} 

This conclusion does not cover Everett-type theories, which reject at the
outset the fundamental idea used here that definite outcomes actually occur.

This extensive discussion of nonlocality  buttresses
the critical assumption of the objective interpretation von Neumann's 
formulation of quantum theory that is being developed here, namely 
the assumption that there is a preferred set of successive 
instants ``now'' associated with the evolving objective quantum state 
of the universe, and that the reduction process acts instantly along
these surfaces. This assumption is completely compatible with the relativistic
covariance of all the {\it  predictions} of relativistic quantum field
theory: that is demonstrated by the Tomonaga-Schwinger formulation of
relativistic quantum field theory [13,14]. 
 
{\bf The Physical World as Information}

Von Neumann quantum theory is designed to yield all the predictions of
Copenhagen quantum theory. It must therefore encompass the 
increments of knowledge that Copenhagen quantum theory makes predictions about.
Von Neumann's theory is, in fact, essentially a theory of the interaction 
of these subjective realities with an evolving objective physical universe.

Von Neumann makes clear the fact that he is trying to tie together the 
subjective perceptual and objective physical aspects of nature:
``it is inherently entirely correct that the measurement  or related
process of subjective perception is a new entity relative to the physical
environment and is not reducible to the latter. Indeed, subjective
perception leads to the intellectual inner life of the individual...''p.418;
``experience only makes statements of the following type: an observer
has made a certain (subjective) observation; and never any like this:
a physical quantity has a certain value.'' p.420:
In the final stage of his analysis he divides the world into parts
I, II, and III, where part ``I was everything up to the retina of the
observer, II was his retina, nerve tracts and brain, and III his
abstract `ego' ''. Clearly, his ``abstract ego'' involves his
consciousness. Von Neumann's formulation of quantum theory develops the 
dynamics of the interaction between these three parts.

The evolution of the physical universe involves three related processes. 
The first is the deterministic evolution of the state of the physical 
universe. It is controlled by the Schroedinger equation of relativistic 
quantum field theory. This process is a local dynamical process, with all
the causal connections arising solely from interactions between neighboring
localized microscopic elements. However, this local process holds only during 
the intervals between quantum events.

Each of these quantum events involves two other processes. The first
is a choice of a Yes-No question by the mind-brain system. The second 
of these two processes is a choice by Nature of an answer, either Yes or No, 
to that question. This second choice is partially free: it is a random
choice, subject to the statistical rules of quantum theory. The first choice is
the analog in von Neumann theory of an essential process in Copenhagen 
quantum theory, namely the free choice made by the experimenter as to which 
aspect of nature is going to be probe. This choice of which aspect of nature
is going to be probed, i.e., of which specific question is going to be put 
to nature, is an essential element of quantum theory: the quantum statistical
rules cannot be applied until, and unless, some specific question is first 
selected.

In Copenhagen quantum theory this choice is made by an experimenter, and
this experimenter lies outside the system governed by the quantum rules. 
This feature of Copenhagen quantum theory is not altered in the transition 
to von Neumann quantum theory:
choice {\it by an individual}, of which question will be put to nature,
is not controlled by any rules that are  known or understood within
contempory physics. This choice on the part of the mind-brain-body system that
constitutes the individual is, in this specific sense, a free choice: it is not
governed by the physical laws of contemporary physics (i.e., quantum
theory). This freedom constitutes a  logical ``gap'' in the dynamical rules
of quantum theory.

Only Yes-No questions are permitted: all other possibilities can be 
reduced to these. Thus each answer, Yes or No, injects one ``bit'' of 
information into the quantum universe. These bits of information are 
stored in the evolving objective quantum state of the universe, which is
a compendium of these bits of information. But it evolves in
accordance with the laws of atomic physics. Thus the quantum state
has an ontological character that is in part matter like, since it
is expressed in terms of the variables of atomic physics, and evolves
between events under the control of the laws of atomic physics. 
However, each event injects the information associated with a subjective
perception by some observing system into the objective state of the
universe. 

This conceptualization of natural process arises not from
some preconceived speculative intuition, but directly from an
examination of the mathematical structure injected into science by our
study of the structure of the relationships between
our experiences. The quantum state state of the universe is thus rooted 
in atomic properties, yet is an informational structure that interacts with, 
and carries into the future, the informational content of each mental event. 
This state has effective causal efficacy because it controls, via 
statistical laws, the propensities for the occurrence of subsequent events.  

Once the physical world is understood in this way, as an atomistically stored 
compendium of locally efficacious bits of information, the instantaneous
transfers of information along the preferred surfaces ``now'' can be
understood to be changes, not in just human knowledge, as in the 
Copenhagen interpretation, but in an absolute state of objective 
information.

{\bf Mind-Brain Interaction and Decoherence}

Von Neumann quantum theory is essentially a theory  of the interaction 
between the evolving objective state of the physical universe and 
a sequence of mental events, each of which is associated with a localized
individual system.  The theory specifies the general form of the 
interaction between subjective knowings associated with individual
physical systems and the physical states of those systems.
The mathematical structure automatically ensures that when the 
state of the individual physical system associated with a mental event 
is brought into alignment with the content of that mental event the 
entire universe is simultaneously brought into alignment with that mental 
content. No special arrangement is needed to produce this key result: 
it is an unavoidable consequence of the quantum entanglements that are
built into the mathematical structure. 

An essential feature of quantum brain dynamics is the strong action of the 
environment upon the brain. This action creates a powerful tendency for 
the brain to transform almost instantly [21] into an ensemble of components,
each of which is very similar to an {\it entire classically-described brain}.
I assume that this transformation does indeed occur, and exploit it in two
important ways. First, this close connection to classical physics makes the 
dynamics easy to describe: classical language and imagery can be 
used to describe in familar terms how the brain behaves.
Second, this description in familar classical terms makes it easy to
identify the important ways in which the actual behaviour differs from 
what classical physics would predict.

A key micro-property of the human brain pertains to the migration of calcium 
ions from the micro-channels through which these ions enter the interior
of nerve terminals to the sites where they trigger the release the
contents of a vesicle of neuro-transmitter. The quantum mechanical rules 
entail [22] that each release of the contents of a vesicle of neurotransmitter
generates a quantum splitting of the brain into different classically 
describable components, or branches.  Evolutionary considerations entail 
that the brain must keep the brain-body functioning in a coordinated way 
and, more specifically, must plan and effectuate, in any normally 
encountered situation, a single coherent course of action that meets 
the needs of that individual. But due to the quantum splitting mentioned 
above, the quantum brain will tend to decompose into components that 
specify alternative possible courses of action. Thus the purely mechanical 
evolution of the state of the brain in accordance with the Schroedinger 
equation will normally causes the brain to evolve into a growing ensemble 
of alternative branches, each of which is essentially 
{\it an entire classically described brain that specifies a possible
plan of action.} 

This ensemble that constitutes the quantum brain is mathematically similar 
to an ensemble that occurs in a classical treatment when  one takes into 
account the uncertainties in our knowledge of the intitial conditions of 
the particles and fields that constitute the classical representation of 
a brain. This close connection between what quantum theory gives and what 
classical physics gives is the basic reason why von Neumann quantum 
theory is able to produce all of the correct predictions of classical physics. 
To unearth specific differences caused by quantum effects one can start 
from this similarity at the lowest-order approximation, which yields 
the classical results, but then dig deeper.

{\bf The Passive and Active Roles of Mind}

The founders of quantum theory recognized that the mathematical
structure of quantum theory is naturally suited for, and seems
to require, bringing into the dynamical equations two separate aspects of 
the interaction between the physical universe and the minds of the 
experimenter/observers. The first of these two aspects is the
role of the experimenter in choosing what to attend to; which aspect of
nature he wants to probe; which question he wants to ask about the
physical world. This is the active role of mind. The second aspect is 
the recognition, or coming to know, the answer that nature returns. 
This is the passive role of mind.
 
{\bf The Active Physical Counterpart to the  Passive Mental Event}

I have mention the Schroedinger evolution of the state $S(t)$ of the
universe. The second part of the orthodox quantum dynamics consists of an 
event that {\it discards} from the ensemble of quasi-classical elements
mentioned above those elements that are incompatible with the answer that 
nature returns. This reduction of the prior ensemble of elements, which
constitute the quantum mechanical representation of the brain, to the 
subensemble compatible with the ``outcome of the query'' is analogous to 
what happens in classical statistical mechanics when new information about 
the physical system is obtained. However, in the quantum case one must 
{\it in principle} regard the {\it entire ensemble of classically described
brains} as real, because interference between the different elements are 
in principle possible.

Each quantum event consists, then of a pair of events, one physical, the other
mental. The physical event reduces the initial ensemble that 
constitutes the brain prior to the event to the subensemble consisting 
of those branches that are compatible with the informational content of 
the associated mental event.

This dynamical connection means that, during an interval of conscious thinking,
the brain changes by an alternation between two processes. The first  
is the generation, by a local deterministic mechanical rule, of an expanding
profusion of alternative possible branches, with each branch corresponding 
to {\it an entire classically describable brain embodying some specific 
possible course of action.} The quantum brain is the {\it entire 
ensemble} of these separate, but equally real, quasi-classical branches. 
The second process involves an event that has both physical and mental 
aspects. The physical aspect, or  event, chops off all branches that are 
incompatible with the associated mental aspect, or event. For example, if the 
mental event is the experiencing of some feature of the physical world, 
then the associated physical event would be the updating of the brain's 
representation of that aspect of the physical world. This updating of the 
(quantum) brain is achieved by {\it discarding} from the ensemble of 
quasi-classical brain states all those branches in which the brain's 
representation of the physical world is incompatible with the information
content of the mental event. 

This connection is similar to a functionalist account of consciouness.
But here it is expressed in terms of a dynamical interaction that is
demanded by the requirement that the objective formulation of the theory 
yield the same predictions about connections between our conscious experiences 
that the empirically validated Copenhagen quantum theory gives. 
The interaction is the exact expression of the basic dynamical rule of 
quantum theory, which is the stipulation that each increment in 
knowledge is associated with a {\it reduction of the quantum state} 
to one that is compatible with the new knowledge. 

The quantum brain is an ensemble of quasi-classical components. As just
noted, this structure is similar to something that occurs in classical 
statistical mechanics, namely a ``classical statistical ensemble.'' 
But a classical statistical ensemble, though structurally
similar to a quantum brain, is fundamentally a different kind of thing. 
It is a representation of a set of truly distinct possibilities,
only one of which is real. A classical statistical ensemble is used when 
a person does not know which of the conceivable possibilities is real, 
but can assign a `probability' to each possibility. In contrast, 
{\it all} of the elements of the ensemble that constitute a 
quantum brain are equally real: no choice has yet been made among them,
Consequently, and this is the key point, entire ensemble acts as a whole in 
the determination of the upcoming mind-brain event.

Each thought is associated with the actualization of some macroscopic 
quasi-stable features of the brain. Thus the reduction event is 
a macroscopic happening. Moreover, this event involves, dynamically, the 
entire ensemble of quasi-classical brain states. In the corresponding 
classical model each element of the ensemble evolves independently, 
in accordance with a micro-local law of motion that involves just that one 
branch alone. Thus there are basic dynamical differences between the quantum 
and classical models, and the consequences of these dynamical differences 
need to be studied in order to exhibit the quantum effects.
 
The only freedom in the theory---insofar as we leave Nature's choices
alone---is the choice made by the individual about {\it which} question 
it will ask next, and {\it when} it will ask it. These are the only inputs 
of mind to the dynamics of the brain. This severe restriction on the role of 
mind is what gives the theory its predictive power. Without this restriction
mind could be free to do anything, and the theory would have no consequences.

Asking a question about something is closely connected to focussing 
one's attention on it. Attending to something is the act of directing 
one's mental power to some task. This task might be to update one's 
representation of some feature of the surrounding world, or to plan or execute 
some other sort of mental or physical action.  

The key question is then: Can this freedom merely to choose
which question is asked, and when it is asked, lead to any 
{\it statistical} influence of mind on the behaviour of the brain,
where a `statistcal' influence is a influence on values obtained by 
averaging over the properly weighted possibilities. 

The answer is Yes!

{\bf The Quantum Zeno Effect}

There is an important and well studied effect in quantum theory
that depends on the timings of the reduction events arising from
the queries put to nature. It is called the Quantum Zeno Effect. It is not
diminished by interaction with the environment [23].  

The effect is simple. If the {\it same} question is put to nature 
sufficiently rapidly and the initial answer is Yes, then any noise-induced 
diffusion, or force-induced motion, of the system away from the subensemble 
where the answer is `Yes' will be suppressed: the system  will tend to be 
confined to the subensemble where the answer is `Yes'. The effect is sometimes
called the ``watched pot'' effect: according to the old adage
``A watched pot never boils''; just looking at it keeps it from changing.
Also, a state can be pulled along in some direction by posing a rapid sequence 
of questions that change sufficiently slowly over time. In short, according to 
the dynamical laws of quantum mechanics, the freedom to choose which questions 
are put to nature, and when they are asked, allows mind to influence  
the behaviour of the brain.  

A person is aware of almost none of the processing that is going on in his
brain: unconscious brain action does almost everthing. So it would be both
unrealistic and theoretically unfeasible to give mind unbridled freedom:
the questions posed by mind ought to be determined in large measure by brain.

What freedom is given to man? 

According to this theory, the freedom given
to {\it Nature} is simply to provide a Yes or No answer to a question posed by 
a subsystem. It seems reasonable to restrict in a similar way the choice 
given to a human mind. 

{\bf A Simple Dynamical Model}

It is easy to construct a simple dynamical model in which the brain 
does most of the work, in a mechanical way, and the mind, by means of 
choices between `Yes' or `No' options, merely gives top-level quidance.

Let $\{P\}$ be the set of projection operator that act only on the
brain-body of the individual and that correspond to possible mental
events of the individual.  Let $P(t)$ be the $P$ in $\{P\}$ that
maximizes $Tr P S(t)$, where $S(t)$ is the state of the universe
at time $t$. This $P(t)$ represents the ``best possible'' question
that could be asked by the individual at time $t$. Let the question 
associated  with $P(t)$ be posed if $P(t)$ reaches a local maximum.
If nature returns the answer `Yes' then the mental event associated
with $P(t)$ occurs. Mental control comes in only through the option
to rapidly pose this same question repeatedly, thus activating the
Quantum Zeno Effect, which will tend to keep the state of the 
brain focussed on the plan of action specified by $P$.

The Quantum  Zeno Effect will not  freeze up the brain completely.
It merely keeps the state of the brain in the subspace where attention 
is focussed on pursuing the plan of action specified by $P$. 

In this model the brain does practically everything, but mind, by means
of the limited effect of consenting to the rapid re-posing of the question 
already constructed and briefly presented by brain, can influence brain 
activity by causing this activity to stay focussed on the presented
course of action.

\noindent {\bf Agreement with Claims of William James}

Does this theory explain anything?

Essentially this  model was already in place [24] when a colleague,
Dr. Jeffrey Schwartz, brought to my attention
some passages from ``Psychology: The Briefer Course'', written by William 
James [25]. In the final section of the chapter on Attention James writes:

``I have spoken as if our attention were wholly 
determined by neural conditions. I believe that the array of {\it things}
we can attend to is so determined. No object can {\it catch} our attention
except by the neural machinery. But the {\it amount} of the attention which
an object receives after it has caught our attention is another question.
It often takes effort to keep mind upon it. We feel that we can make more 
or less of the effort as we choose. If this feeling be not deceptive, 
if our effort be a spiritual force, and an indeterminate one, then of 
course it contributes coequally with the cerebral conditions to the result.
Though it introduce no new idea, it will deepen and prolong the stay in 
consciousness of innumerable ideas which else would fade more quickly
away. The delay thus gained might not be more than a second in duration---
but that second may be critical; for in the rising and falling 
considerations in the mind, where two associated systems of them are
nearly in equilibrium it is often a matter of but a second more or 
less of attention at the outset, whether one system shall gain force to
occupy the field and develop itself and exclude the other, or be excluded 
itself by the other. When developed it may make us act, and that act may 
seal our doom. When we come to the chapter on the Will we shall see that 
the whole drama of the voluntary life hinges on the attention, slightly 
more or slightly less, which rival motor ideas may receive. ...''  
 
In the chapter on Will, in the
section entitled ``Volitional effort is effort of  attention'' 
James writes:

``Thus  we find that {\it we reach the  heart  of our inquiry  into volition
when we ask by what process is it that the thought of any given action
comes to prevail stably in the mind.}'' 

and later

``{\it  The essential achievement of the will, in short, when it is most 
`voluntary,'  is to attend to a difficult  object and hold it fast before
the  mind.   ...  Effort of attention is  thus the essential phenomenon
of will.''}

Still  later, James says:

{\it  ``Consent to the idea's undivided presence, this is effort's sole 
achievement.''} ...``Everywhere, then, the function  of effort is the same:
to keep affirming and adopting the thought  which,  if left to  itself, would 
slip away.''
  
This description of the effect of mind on the course of mind-brain process 
is remarkably in line with what had been proposed independently from purely 
theoretical consideration of the quantum physics of this process. 
The connections claimed by James are explained of the basis of the same 
dynamical principles that had  been introduced by physicists explain 
atomic phenomena. Thus the whole range of science, from atomic physics 
to mind-brain dynamics, is brought together in a single rationally 
coherent theory of an evolving cosmos that consists of a physical reality
that represents information, interacting via the laws of atomic physics
with the closely related, but differently constituted, mental aspects of 
nature.

{\bf Agreement With Recent Work on Attention}

Much experimental work on attention and effort has occurred 
since the time of William James. That work has been 
hampered by the nonexistence of any putative physical theory 
that purports to explain how our conscious experiences 
influence activities in our brains. The behaviourist approach,
which dominated psychological during the first half of the
twentieth century, and which essentially abolished, in this field,
not only the use of introspective data but also the very concept of 
consciousness, was surely motivated in part by the fact that
consciousness had no natural place within the framework of classical 
physical theory. According to the principles of classical physical theory, 
consciousness makes no difference in behavior: all behavior is,  
determined by microscopic causation without ever acknowledging the 
existence of consciousness. Thus philosophers who accepted the ideas
of classical physics were driven to conclude that conscious experiences 
were either {\it identical} to corresponding classically describable 
activities of the brain, or were ``emergent'' properties. The first idea, 
the identity theory of mind, seems impossible to reconcile with the fact 
that according to the classical principles the brain is an assembly of 
local elements behaving in accordance with the local laws of classical 
physics, and that all higher-order dynamical properties are just 
re-expressions of the local causal links between the local elements.
But the existence of ``feelings'' and other conscious experiences is not
just a re-expression of the causal links described by the
principles of classical physical theory. And any ``emergent'' property
that emerges from a system whose behavior is completely specified by the
classical principles is only {\it trivially emergent}, in the same sense 
as is the {\it wheelness} often cited by Roger Sperry: ``wheelness'' 
did not exist in the physical world before wheels, and it exerts 
top-down causation, via the causal links specified by the classical 
principles. But the emergence of ``wheelness'' is not analogous to the 
emergence of ``consciousness'': the existence of the defining 
characteristics of the ``wheelness'' of a wheel follows rationally from a 
classical physics model of a wheel, but the existence of the defining 
experiential characteristics of the ``consciousness''  a brain 
does not follow rationally from a classical physics model of the brain.

The failure of the behaviourist programs led to the rehabilitation
of ``attention'' during the early fifties, and many hundreds 
of experiments have been performed during the past fifty years
for the purpose of investigating empirically those aspects 
of human behaviour that we ordinarily link to our consciousness. 

Harold Pashler's 1998 book ``The Psychology of Attention'' [26]
describes a great deal of this empirical work, and also the 
intertwined theoretical efforts to understand the nature
of an information-processing system that could account for 
the intricate details of the objective data. Two key concepts 
are the notions of a processing ``Capacity'' and of ``Attention''. 
The latter is associated with an internally directed 
{\it selection} between different possible allocations 
of the available processing ``Capacity''. A
third concept is ''Effort'', which is linked to
incentives, and to reports by subjects of ``trying harder''.

Pashler organizes his discussion by separating perceptual
processing from postperceptual processing. The former covers 
processing that, first of all, identifies such basic physical
properties of stimuli as location, color, loudness, and pitch, 
and, secondly, identifies stimuli in terms of categories of meaning.
The postperceptual process covers the tasks of producing motor actions
and cognitive action beyond mere categorical identification.
Pashler emphasizes [p. 33] that ``the empirical findings of attention 
studies specifically argue for a distinction between perceptual 
limitations and more central limitations involved in thought 
and the planning of action.'' The existence of these two different
processes, with different characteristics, is a principal theme of
Pashler's book [p. 33, 263, 293, 317, 404].

In the quantum theory of mind-brain being described here there 
are two separate processes. First, there is the unconscious 
mechanical brain process governed by the Schroedinger equation. 
As discussed in ref. 22, this brain processing involves  
dynamical units that are represented by complex patterns of neural 
activity (or, more generally, of brain activity) that are ``facilitated'' 
by use, and such that each unit tends to be activated as a whole by 
the activation of several of its parts: this explains the development of brain
process through ''association''. The brain evolves mechanically by 
the dynamical interplay of these dynamic units, and by feed-back loops 
that strenghthen or weaken appropriate input channels.

Each individual quasi-classical element of the ensemble of alternative
possible brain states that constitutes the quantum brain creates, on the 
basis of clues, or cues, coming from various sources, a plan for a possible 
coherent course of action. Quantum uncertainties entail that a host of 
different possibilities will emerge, and hence that the quantum brain will
evolve into a set of component classically describable brains representing 
different possible courses of action. [See ref. 22.] This mechanical
phase of the processing already involves some selectivity, because the various 
input clues contribute either more or less to the evolving brain process 
according to the degree to which these inputs activate, via associations,
the patterns that survive and turn into the plan of action.

This conception of brain dynamics seems to accommodate all of 
the perceptual aspects of the data described by Pashler. But it is 
the high-level processing, which is more closely linked to our active 
mentally controlled conscious thinking, that is of prime interest here. 
The data pertaining to this second process is the focus of part II of 
Pashler's book.

Mental intervention has, according to the quantum-physics-based theory 
described here, several distinctive characteristics. It consists of a 
sequence of discrete events each of which consents to an 
integrated course of action presented by brain. The rapidity of 
these events can be increased with effort.  Effort-induced speed-up 
of the rate of occurrence of these events can, by means of the 
quantum Zeno effect, keep attention focussed on a task. Between 100 and 
300 msec of consent seem to be needed to fix a plan of action. Effort can, 
by increasing the number of events per second, increase the mental 
input into brain activity. Each conscious event picks out from the
multitude of quasi-classical possibilities that comprise the quantum brain 
the subensemble that is compatible with the conscious experience. 

The correspondence between the mental event and the associated physical 
event is this: the physical event reduces the prior physical ensemble 
of alternative possibilities to the subensemble compatible with the 
mental event. This connection will be recognized as the core interpretive 
postulate of Copenhagen quantum theory: the physical event reduces
the prior state of the observed system to the part of it that is compatible 
with the experience of the observer.

Examination of Pashler's book shows that this quantum-physics-based theory
accommodates naturally  all of the complex structural features 
of the empirical data that he describes. He emphasizes [p. 33] a 
specific finding: strong empirical evidence for what he calls a central 
processing bottleneck associated with the attentive selection of a motor 
action. This kind of bottleneck is what the quantum-physics-based theory 
predicts: the bottleneck is precisely the single linear sequence of mind-brain 
quantum events that von Neumann-Wigner quantum theory is built upon. 

Pashler [p. 279] describes four empirical signatures for this kind of 
bottleneck, and describes the experimental confirmation of each of them. 
Much of part II of Pashler's book is a massing of evidence that  
supports the existence of a central process of this general kind.

This bottleneck is not automatic within classical physics. A classical 
model could easily produce simultaneously two responses in different 
modalities, say vocal and manual, to two different stimuli arriving via 
two different modalities, say auditory and tactile: the two processes 
could proceed via dynamically independent routes. Pashler [p. 308]
notes that the bottleneck is undiminished in split-brain 
patients performing two tasks that, at the level of input and output, 
seem to be confined to different hemispheres.

Pashler states [p. 293] ``The conclusion that there is a central 
bottleneck in the selection of action should not be confused with
the ... debate (about perceptual-level process) described in chapter 1.
The finding that people seem unable to select two responses at the same 
time does not dispute the fact that they also have limitations in perceptual
processing...''. I have already mentioned the independent selectivity
injected into brain dynamics by the purely mechanical part of the 
quantum mind-brain process. 

The queuing effect for the mind-controlled motor responses does not 
exclude interference between brain processes that are similar
to each other, and hence that use common brain mechanisms. Pashler [p. 297] 
notes this distinction, and says ``the principles governing queuing
seem indifferent to neural overlap of any sort studied so far.'' 
He also cites evidence that suggests that the hypthetical timer of 
brain activity associated with the cerebellum ``is basically independent 
of the central response-selection bottleneck.''[p. 298]

The important point here is that there is in principle, in the quantum
model, an essential dynamical difference between the unconscious processing
carried out by the Schroedinger evolution, which generates via a local
process an expanding collection of classically conceivable possible courses 
of action, and the process associated with the sequence of conscious events 
that constitutes a stream of consciousness. The former are not limited by
the queuing effect, because all of the possibilities develop in parallel, 
whereas the latter do form elements of a single queue. The experiments
cited by Pashler all seem to support this clear prediction of the quantum
approach.

An interesting experiment mentioned by Pashler involves the simultaneous
tasks of doing an IQ test and giving a foot response to a rapidly 
presented sequences of tones of either 2000 or 250 Hz. The subject's mental 
age, as measured by the IQ test, was reduced from adult to 8 years. [p. 299]
This result supports the prediction of quantum theory that the bottleneck 
pertains to both `intelligent' behaviour, which requires conscious processing, 
and selection of motor response.

Another interesting experiment showed that, when performing at maximum
speed, with fixed accuracy, subjects produced responses at the 
same rate whether performing one task or two simultaneously: the 
limited capacity to produce responses can be divided between two 
simultaneously performed tasks. [p. 301]

Pashler also notes [p. 348] that ``Recent results strengthen the case for 
central interference even further, concluding that memory retrieval is subject
to the same discrete processing bottleneck that prevents simultaneous response
selection in two speeded choice tasks.''

In the section on ``Mental Effort'' Pashler reports that 
``incentives to perform especially well lead subjects to improve both 
speed and accuracy'', and that the motivation had ``greater effects
on the more cognitively complex activity''. This is what would be 
expected if incentives lead to effort that produces increased rapidity of 
the events, each of which injects into the physical process, via quantum 
selection and reduction, bits of control information that reflect mental  
evaluation.

Studies of sleep-deprived subjects suggest that in these cases ``effort 
works to counteract low arousal''. If arousal is essentially the rate
of occurrence of conscious events then this result is what the quantum model 
would predict. 

Pashler notes that ``Performing two tasks at the same time, 
for example, almost invariably... produces poorer performance in a task and 
increases ratings in effortfulness.'' And ``Increasing the rate at which 
events occur in experimenter-paced tasks often increases effort ratings 
without affecting performance''. ``Increasing incentives often raises 
workload ratings and performance at the same time.'' All of these 
empirical connections are in line with the general principle that 
effort increases the rate of conscious events, each of which inputs a
mental evaluation and a selection or focussing of a course of 
action, and that this resource can be divided between tasks.

Additional supporting evidence comes from the studies of the effect
of the conscious process upon the storage of information in 
short-term memory. According to the physics-based  theory, the conscious
process merely actualizes a course of action, which then develops
automatically, with perhaps some occasional monitoring. Thus if one 
sets in place the activity of retaining in memory a certain sequence
of stimuli, then this activity can persist undiminished while the 
central processor is engaged in another task. This is what the data
indicate. 

Pashler remarks that
''These conclusions contradict the remarkably widespread assumption
that short-term memory capacity can be equated with, or used as a 
measure of, central resources.''[p.341] In the theory outlined here
short-term memory is stored in patterns of brain activity, whereas 
consciousness is associated with the selection of a subensemble
of quasi-classical states. This distinction seems to account for
the large amount of detailed data that bears on this question of 
the connection of short-term-memory to consciousness. [p.337-341]

Deliberate storage in, or retrieval from, long-term memory requires 
focussed attention, and hence conscious effort. These processes should,
according to the theory, use part of the limited processing capacity, 
and hence be detrimentally affected by a competing task that makes 
sufficient concurrent demands on the central resources. On the other hand,  
``perceptual'' processing that involves conceptual categorization and 
identification without conscious awareness should not interfer with 
tasks that do consume central processing capacity. These expectations
are what the evidence appears to confirm: ``the entirety of...front-end
processing are modality specific and operate independent of the sort of 
single-channel central processing that limits retrieval and the control 
of action. This includes not only perceptual analysis but also storasge
in STM (short term memory) and whatever may feed back to change the 
allocation of perceptual attention itself.'' [p. 353]

Pashler describes a result dating from the nineteenth century:
 mental
exertion reduces the amount of physical force that a person can
apply. He notes that ``This puzzling phenomena remains unexplained.''
[p. 387]. However, it is an automatic consequence of the physics-based theory:
creating physical force by muscle contraction requires an effort 
that opposes the physical tendencies generated by the Schroedinger
equation. This opposing tendency is produced by the quantum Zeno effect,
and is roughly proportional to the number of bits per second of central
processing capacity that is devoted to the task. So if part of this 
processing capacity is directed to another task, then the applied force
will diminish.
 
Pashler speculates on the possibility of a  neurophysiological
explanation of the facts he describes, but notes that the parallel 
versus serial distinction between the two mechanisms leads, in the classical 
neurophysiological approach, to the questions of what makes these two 
mechanisms so different, and what the connection between them is. [p.354-6,
386-7]

After analyzing various possible mechanisms that could cause the central 
bottleneck, Pashler [p.307-8] says ``the question of why this should be 
the case is quite puzzling.'' Thus the fact that this bottleneck, and its
basic properties, follow automatically from the same laws that explain 
the complex empirical evidence in the fields of classical and quantum physics 
means that the theory has significant explanatory power.

Of course, some similar sort of structure could presumably be 
worked into a classical model. But a general theory of all of nature that  
automatically explains a lot of empirical data in a particular field on 
the basis of the general principles is normally judge superior to a 
special theory that is rigged after the fact to explain these data.

It needs to be emphasized that there is at present absolutely no
empirical evidence that favors the classical model over the quantum 
model described above.  The classical model would have to be implemented
as a statistical theory, due to the uncertainties in  the  initial
conditions, and that statistical  model is to first order the
same as the simple quantum model described above.  The quantum
model has the advantage that at least it {\it could} be valid,
whereas the classical model must necessarily fail when quantum
effects become important.   So nothing is  lost by switching to
quantum theory, but a lot is gained.  Psychology and psychiatry gain
the possibility of reconciling with neuroscience the essential 
psychological concept of the ability of our minds to guide our actions.
And psycho-physics gains a dynamical model for the interaction of mind 
and brain. Finally, philosophy of mind is liberated from the dilemma
of having to choosing between ``identity theory'' and the ``emergence'' of a 
causally inert mind: the emergence a mind which lacks the capacity to 
act back on the physical world in the way needed to drive its evolution
hand-in-hand with the evolution of the brain.

\begin{center}
\vspace{.2in}
\noindent {\bf References}
\vspace{.2in}
\end{center}

1. Physics Today, December 1998, p. 9.

2. W. Tittle, J. Brendel, H. Zbinden, and N. Gisin, 
        Phys. Rev. Lett. {\bf 81}, 3563 (1998).

3. W. Tittle, J. Brendel, H. Zbinden, and N. Gisin, 
        Phys. Rev. {\bf A59}, 4150 (1999).

4. N. Bohr, Phys. Rev. {\bf 48}, 696 (1935).

5. P.A.M. Dirac, at 1927 Solvay Conference {\it Electrons et photons:
Rapports et Discussions du cinquieme Conseil de Physique}, Gauthier-Villars, 
Paris, 1928.

6. W. Heisenberg, Daedalus {\bf 87}, 95-108 (1958).

7. A. Einstein, B. Podolsky, and N. Rosen,  Phys. Rev. {\bf 47}, 777 (1935). 

8. N. Bohr, Phys. Rev. {\bf 48}, 696 (1935).

9. A. Einstein, in {\it Albert Einstein: Philosopher-Physicist},
   ed, P. A.  Schilpp,   Tudor, New York, 1951.  p.669.

10. M. Gell-Mann, in {\it The Nature of the Physical Universe: the 1976
    Nobel Conference}, Wiley, New York, 1979, p. 29. 

11. J. von Neumann, {\it Mathematical Foundations of Quantum Mechanics},\\ 
    Princeton University  Press, Princeton, NJ, 1955;\\
    Translation from the 1932 German original. 

12. N. Bohr, Atomic Physics and Human Knowledge, Wiley, New York,
    1958, p.88, p.72.   

13. S. Tomonaga, Progress of Theoretical Physics, {\bf 1},  (1946) 

14. J. Schwinger, Physical Review, {\bf 82}, 914 (1951).

15. G.F. Smoot et. al., Astrophysical Journal {\bf 396}, L1 (1992).

16. J.S. Bell, Physics, {\bf 1}, 195 (1964); and in {\it Speakable and
    Unspeakable in Quantum Mechanics.} Cambridge Univ. Press, (1987) Ch. 4;   
    J. Clauser and A. Shimony, Rep. Prog. Phys. {\bf 41}, 1881 (1978).

17. H. Stapp, Epistemological Letters, June 1978. (Assoc. F Gonseth,
          Case Postal 1081, Bienne Switzerland).

18. A Fine, Phys. Rev. Lett. {\bf 48}, 291 (1982).

19. N. Bohr, ``Atomic Physics and Human Knowledge'', Wiley, New York,
    (1958) p.72

20. L. Hardy, Phys. Rev. Lett. {\bf 71}, 1665 (1993):\\
    A. White, D. F. V. James, P. Eberhard, and P.G. Kwiat,\\
    Physical Review Letters, {\bf 83},  3103  (1999).

21. Max Tegmark, ``The Importance of Quantum Decoherence in Brain
    Process,''  Phys. Rev E, {\bf 61}, 4194-4206 (2000).

22. H.P. Stapp, ``Mind, Matter, and Quantum Mechanics'', 
    Springer-Verlag, New York, Berlin. (1993) p.152, Ch VI.

23. H.P. Stapp, ``The Importance of Quantum Decoherence in Brain
    process'', Lawrence Berkeley  National Laboratory Report LBNL-46871.\\
    quant-ph 0010029

24. H. Stapp, ``Attention, Intention, and Will in Quantum Physics''\\
    in Journal of Consciousness Studies, {\bf 6}, 143 (1999); 

25. Wm. James, ``Psychology: The  Briefer Course'', ed. Gordon Allport,
    University of Notre Dame Press, Notre Dame, IN.  Ch. 4 and Ch. 17

26. Harold Pashler,  ``The Psychology  of Attention'',\\
    MIT Press, Cambridge MA (1998)

\end{document}